\documentstyle[prl,aps,amsfonts,preprint]{revtex}
\input{psfig.sty}
\newcommand{\ar}{\arrowvert}
\newcommand{\ra}{\rangle}
\newcommand{\la}{\langle}
\newcommand{\da}{\dagger}

\begin{document}
\draft
\title{Many-Body Coulomb Gauge
Exotic and Charmed Hybrids}
\author{Felipe J. Llanes-Estrada and Stephen R. Cotanch}
\address{Department of Physics, North Carolina State
University, Raleigh NC  27695-8202 USA}
\date{\today}
\maketitle
\begin{abstract}
Utilizing a QCD Coulomb gauge Hamiltonian with linear confinement 
specified by lattice, we report a relativistic many-body  calculation for 
the light exotic and charmed hybrid mesons. The Hamiltonian  successfully 
describes both quark  and gluon  sectors, with vacuum and  quasiparticle 
properties  generated by a BCS transformation and more elaborate TDA and 
RPA  diagonalizations for the meson ($q\overline{q}$) and glueball 
($gg$)  masses. Hybrids entail a computationally intense relativistic three
quasiparticle ($q\overline{q}g$) calculation with the 9 dimensional
Hamiltonian matrix elements evaluated variationally by Monte Carlo 
techniques. Our new TDA (RPA) spectrum for the nonexotic $1^{--}$ charmed
($c\overline{c}$ and $c\overline{c}g$) system provides an explanation for 
the overpopulation of the observed $J/\psi$ states. For the important
$1^{-+}$ light exotic channel we obtain  hybrid  masses above 2
$GeV$, in broad agreement with lattice  and flux tube models, indicating
that the recently observed resonances at 1.4 and 1.6 $GeV$ are of
different, perhaps four quark, structure.
\end{abstract}

Exotic hybrids, hadrons with
quantum numbers not possible in simple $q\overline{q}$
or $qqq$ quark
models, have been an
elusive, yet signature   prediction of
Quantum Chromodynamics
(QCD).  It was
therefore quite natural that
the
recent observation by the E852
collaboration \cite{exp}
of two exotic $J^{P
C}=1^{-+}$ states with
masses 1.4 and 1.6 $GeV$ would attract widespread
interest.
Since these states
have isospin $I =1$ they can not be glueballs
(oddballs)
and would initially appear to be viable hybrid meson candidates,
especially
since vintage bag model calculations \cite{bag}
predict exotic
excitations with
explicit gluonic degrees of freedom in this mass range.
However,
the detailed structure of these states remains uncertain since
most
contemporary theoretical studies, such as lattice gauge
\cite
{lacock,hybridlat1,Manke,morn}, flux tube \cite{flux,flux2}, QCD sum 
rule
\cite{sum}, perturbative non-relativistic QCD \cite {bram} and
constituent
models \cite{constit}
have focused on heavy quark hybrids.
The only two
modern light quark hybrid
calculations, a lattice gauge
\cite{hybridlat1}
and flux tube
\cite{flux2}, have further
compounded this uncertainty
by
predicting the lightest exotic hybrid mass to be about
$2.0$ $GeV$--
significantly above
the observed $1^{-+}$ states.
Because the bag model
results are rather
dated and lattice calculations
are less accurate for
light quarks due
to extrapolation, it is important to have an additional,
alternative
hybrid prediction.  The purpose of  this Letter is to determine
if these
exotic states can
indeed be interpreted as hybrids within a
relativisitc
many-body constituent approach that has
successfully described
both conventional meson \cite{flesrc,flsc} and 
glueball \cite{ssjc96}
systems.

Our starting point is the QCD Coulomb gauge Hamiltonian
(see
for example ref. \cite{ChristLee}) which we simplify to a  form
amenable
for many body calculations

\begin{eqnarray}
\label{Hamiltonian}
\nonumber
H = \int d {\bf x} \Psi ^{\dagger}
({\bf x})
(-i{\bf \alpha}\!\cdot \! {\nabla} + \beta m )
\Psi({\bf x}) + Tr \int
d{\bf x} (
{\bf \Pi}^a  \!\cdot \!  {\bf \Pi}^a +
{\bf B}_{A}^a  \!\cdot \!
{\bf B}_{A}^a )
\\ - \frac{1}{2} \int d {\bf x} d {\bf y}
\rho^a({\bf
x})V(\arrowvert {\bf x} -
{\bf y} \arrowvert) \rho^a({\bf y})    \ .
\end{eqnarray}
Here  $\Psi$ and  $\bf A$, are the respective quark
and
gluon fields, ${\bf B}_{A}^a =  \nabla \times
{\bf A}^a $, and
$\rho^a
= \Psi^{\dagger} T^a \Psi + f^{abc} {\bf A}^b
\cdot {\bf \Pi}^c$
is the
quark plus gluon color density.
The current quark mass, $m$, is assigned
the
values,
$m_u = m_d = 5 \ MeV$ and $m_c = 1200 \ MeV$ for the $u, d$ and
$c$
flavors, respectively. Confinement and leading canonical interactions
are
represented by the instantaneous potential,  $V= -
\frac{\alpha_s}{r}
+\sigma r$, with $\alpha_s = .2$,
and,  $\sigma=0.135$
$GeV^2$, as
determined by the string tension from
lattice and Regge fits.
We also use a
cut-off parameter $\Lambda=4-5 \  GeV$ to regularize
the logarithmic
divergent term in the mass gap
equation. The model
parameters
$\sigma$,
$\alpha_s$ and $\Lambda$ are
commensurate with our
previous pure  quark
\cite{flesrc,flsc} and
gluon
\cite{ssjc96}
applications which produced
reasonable hadronic
descriptions
including the  Regge trajectory slopes
for
the mesons (e.g.
$\rho$ tower) and glueballs (pomeron)
\cite{lisbon}.

Next we proceed to
the many-body diagonalizations
but
first perform a canonical transformation
(BCS rotation) to a
new
quasiparticle basis
\begin{eqnarray}
\label{BVgluerotation}
\alpha^a_i({\bf{k}}) = \cosh \Theta_k
\,
a^a_i({\bf{k}}) +
\sinh \Theta_k
\, a_i^{a\da}(-{\bf{k}})
\\
\nonumber
B_{c \lambda }({\bf{k}}) = \cos
\frac{\theta_k}{2} b_{c \lambda
}({\bf{k}})
 -\lambda
\sin
\frac{\theta_k}{2} d_{c \lambda
}^{\dagger}({-{\bf{k}}})
\\
\nonumber
D_{c \lambda }(-{\bf{k}}) = \cos
\frac{\theta_k}{2} d_{c \lambda
}(-{\bf{k}})
 +\lambda\sin
\frac{\theta_k}{2} b^{\dagger}_{c \lambda
}({\bf{k}})  \
,
\end{eqnarray}
where $\Theta_k$, $ \theta_k/2 $ are the
BCS angles,
further specified below,
and
$a (\alpha), b(B)$ and
$d(D)$ are
bare
(dressed) gluon, quark and antiquark Fock
operators,
respectively.
The
indices $a = 1,2...8$ and $c = 1, 2, 3$
denote color  while $\lambda$
represents spin projection.
The new field
expansions
are \begin{eqnarray}
\label{colorfields2}
A^a_i({\bf{x}})
= \int
\frac{d{\bf{k}}}{(2\pi)^3}
\frac{1}{\sqrt{2\omega_k}}[\alpha^a_i({\bf{k}} )
+ \alpha^{a\dag}_i(-{\bf{k}})]
e^{i{\bf{k}}\cdot {\bf
{x}}}
\\
\Pi_i^a({\bf{x}}) = -i \int
\frac{d{\bf{k}}}{(2\pi)^3}
\sqrt{\frac{\omega_k}{2}} [\alpha_i^a({\bf{k}})-
\alpha^{a\dag}_i(-{\bf{k}})]e^{i{\bf{k}}\cdot
{\bf{x}}} \ ,
\end{eqnarray}
for the gluon fields and

\begin{equation}
\Psi({\bf{x}})=\sum_{c
\lambda}
\int \frac{d{\bf{k}}}{(2\pi)^3}
\left[U_{c\lambda}({\bf{k}})B_{c\lambda}({\bf{k}})
+
V_{c\lambda}(-{\bf{k}})D^{\dagger}_{c\lambda}(-{\bf{k}}) \right] e^{i{\bf{k}}
\cdot {\bf{x}}} \ , \end{equation}
for the fermion field. The
rotated Dirac spinors are given in terms of the
Pauli spinors, $\chi$,
\begin{equation} U_{c \lambda} ({\bf{k}}) =
\frac{1}{\sqrt{2}}
\left[ \begin{array}{c}
\sqrt{1+\sin\phi_{k}}\  \chi_{c
\lambda}    \\
\sqrt{1-\sin \phi_k}
\, \, {\bf{\sigma}}  \cdot \hat{\bf
{k}} \, \chi_{c
\lambda} \end{array}
\right] \ ,
\
{\it  V}_{c
\lambda}({\bf { k}}) =
\frac{1}{\sqrt{2}} \left[
\begin{array}{c}
-\sqrt{1-\sin \phi_k} \,
{\bf{\sigma}}\cdot  \hat {\bf
{k}} \, \chi_{c
\lambda} \\
\sqrt{1+\sin
\phi_k} \  \chi_{c
\lambda}\end{array} \right] \ . \end{equation}

We then find an improved,
nontrivial vacuum,
$\arrowvert \Omega \rangle$,  by minimizing the ground
state expectation value of the Hamiltonian
variationally with respect
to the BCS angles.  Actually, the specific
variational parameters are the
quark gap angle, $\phi_k$, related to the
BCS angle by $tan (\phi_k - \theta_k) = m /k$,
and the gluon self-energy,
$\omega_k$, satisfying
$\omega_k = k
e^{-2\Theta_k}$.
This generates a
quark and gluon gap
equation (equivalent to the
Schwinger-Dyson equation)
yielding  mass gaps
of about 100
$MeV$ for the $u/d$ quarks and 800
$MeV$
for the  gluon.  The
BCS vacuum contains quark and gluon
condensates
(Cooper
pairs) in
reasonable agreement with QCD sum rules. For
more complete
details consult
refs.
\cite{flesrc,flsc,ssjc96}.

We find the two body sector is
adequately described by the
Tamm-Dancoff approximation (TDA) in which a
glueball is represented by Fock states
$g^\dagger g^\dagger \arrowvert
\Omega \rangle$ and mesons
by $q^\dagger \overline{q}^\dagger \arrowvert
\Omega \rangle$.
The notable
exception is the light pseudoscalar sector, where a
more sophisticated,
collective approximation (Random Phase approximation or
RPA) is needed to correctly reproduce the Goldstone boson
nature of the pion due to
spontaneous chiral symmetry breaking by our BCS vacuum.
This is also further documented in refs.
\cite{flesrc,flsc}.

\begin{figure}
\psfig{figure=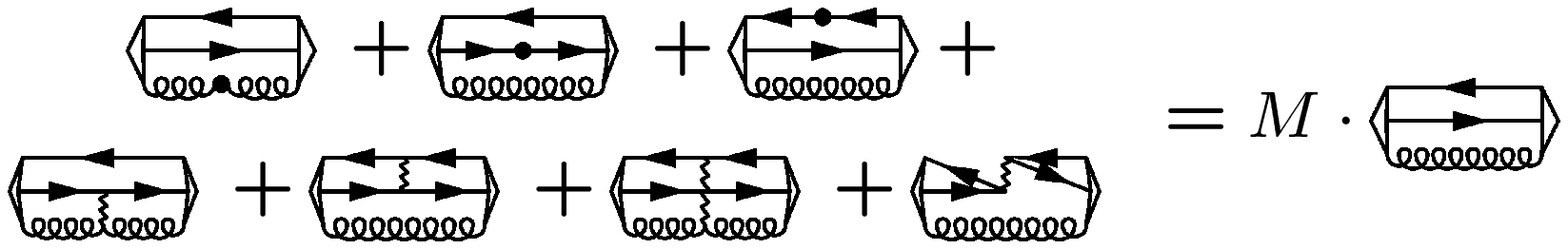,width=7.0in,height=8.5in}
\vspace{-4in}
\caption{One (dot) and two-body
(waves) TDA matrix element for hybrid mesons with one
constituent gluon.
Note the $q\overline{q}$ annihilation two-body
diagram.}
\label{Matelement}
\end{figure}

Finally we formulate the hybrid meson as $
\arrowvert hybrid\rangle =
q^{\dagger}\overline{q}^{\dagger}g^{\dagger}\arrowvert
\Omega
\rangle
\equiv [[q^\dagger
\bigotimes \overline{q}^\dagger]_8
\bigotimes g^\dagger]_0 \arrowvert \Omega \rangle$,
now involving quark
color octet states.
The resulting TDA equation for the hybrid mass $M$
is \begin{equation}
\langle hybrid
\arrowvert [H,q^{\dagger}\overline{q}^{\dagger}g^{\dagger}]
\arrowvert \Omega \rangle = M \langle  hybrid
\arrowvert
q^{\dagger}\overline{q}^{\dagger}g^{\dagger}
\arrowvert
\Omega
\rangle \ .
\end{equation}
This projects the hybrid meson wave
equation,
which is pictorially represented in Fig. \ref{Matelement}, onto
the three
body Fock basis. Unlike our pion
application, the $q\overline{q}$
pair is now in
a  color octet and the TDA is sufficient
since the
Hamiltonian, as well as
exact QCD, does  not conserve the chiral
color octet current (we
specifically calculated no  difference in the more
elaborate, chiral symmetry preserving hybrid RPA calculation, see below).
The relevant
angular momenta ($am$) are  the $q,\overline{q}$ spins coupled
to
an intermediate $S$, and the gluon spin  with its orbital
$am$ $L_+$
(with respect to the $q\overline{q}$ cm) coupled to
intermediate $l$.
Coupling
$l$ with $L_-$ (the orbital $q\overline{q} \ am$) yields
$L$ which
combines with $S$ giving the total $am \ J$.
The complete wavefunction in
the hybrid cm has form

\begin{eqnarray}
F^{J P C}_{\lambda_g
\lambda_q
\lambda_{\overline{q}}}({\bf q}_+,{\bf q}_-)
= \sum_{l L_- L_+ L
S m_+
m_-}
F^{ J P C}_{l L_- L_+  LS}(\ar {\bf q}_+\ar,\ar {\bf q}_-\ar)
\,
\,
Y^{m_+}_{L_+}(\hat{{\bf
{q}}}_+)
\, Y^{m_-}_{L_-}(\hat{{\bf q}}_-)
\\
\nonumber
(-1)^{\lambda_g} \la L_+ m_+ 1 -\lambda_g \ar l m_l \ra \la
L_-
m_- l m_l
\ar L m_L
\ra \la
\frac{1}{2}
\lambda_q
\frac{1}{2}
-\lambda_{\overline{q}} \ar S m_S \ra
(-1)^{\frac{1}{2}
-
\lambda_{\overline{q}}} \la L m_L S m_S \ar J m_J \ra
\
,
\end{eqnarray}
where ${\bf q}_-$ and ${\bf q}_+$ are the
respective
relative momentum of the
$q\overline{q}$ pair and gluon (with
respect to
the pair cm).
We then impose the transversality condition,
$\hat{\bf
{k}}\cdot
{\bf{\alpha}}({\bf{k}}) =0$, from the Coulomb gauge
constraint
which eliminates
states with $L_+=1$ and $l=0$.
For pure $S$
waves the
lightest hybrid states will then have
$J^{PC}
=$
$1^{+-}$,
$0^{++}$,
$1^{++}$ and
$2^{++}$.  These are nonexotic
states
which will mix with conventional
mesons and hinder
hybrid
identification.
For exotic states one $P$ wave is necessary and
we
calculate the
lightest corresponds to
$L_+ = 1$ since the $L_{-} =
1$
excitation is energetically more expensive
due to
quark repulsion in
the
octet channel. This  generates the exotic
states
$1^{-+}$,
$3^{-+}$
and
$0^{--}$.

\begin
{figure}
\psfig{figure=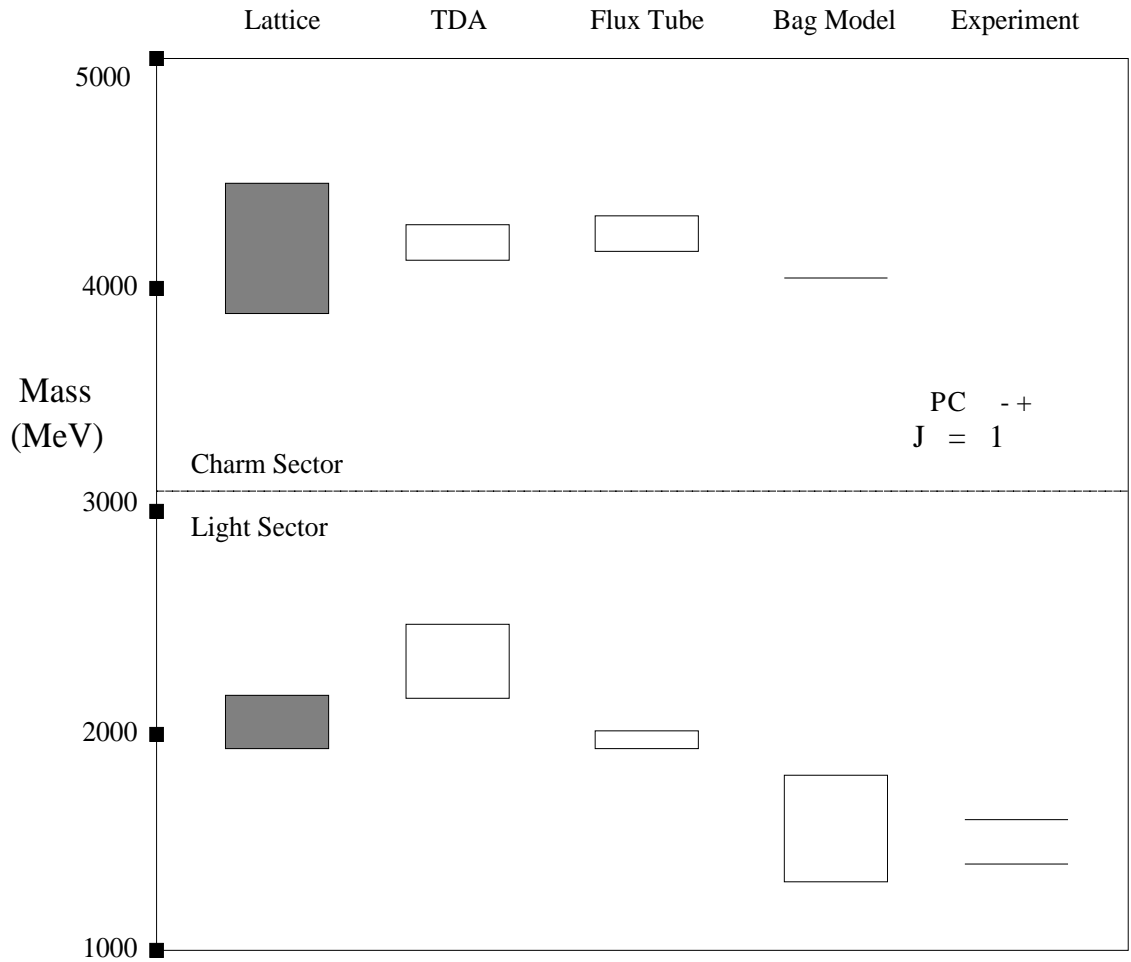,width=6in,height=5in}
\vspace{0.5in}
\caption{Comparison
of exotic
$1^{-+}$ $u/d$ and $c$ hybrid masses with
alternative theories
and data.}
\label{spectrum}
\end{figure}

\begin{figure}
\psfig{figure=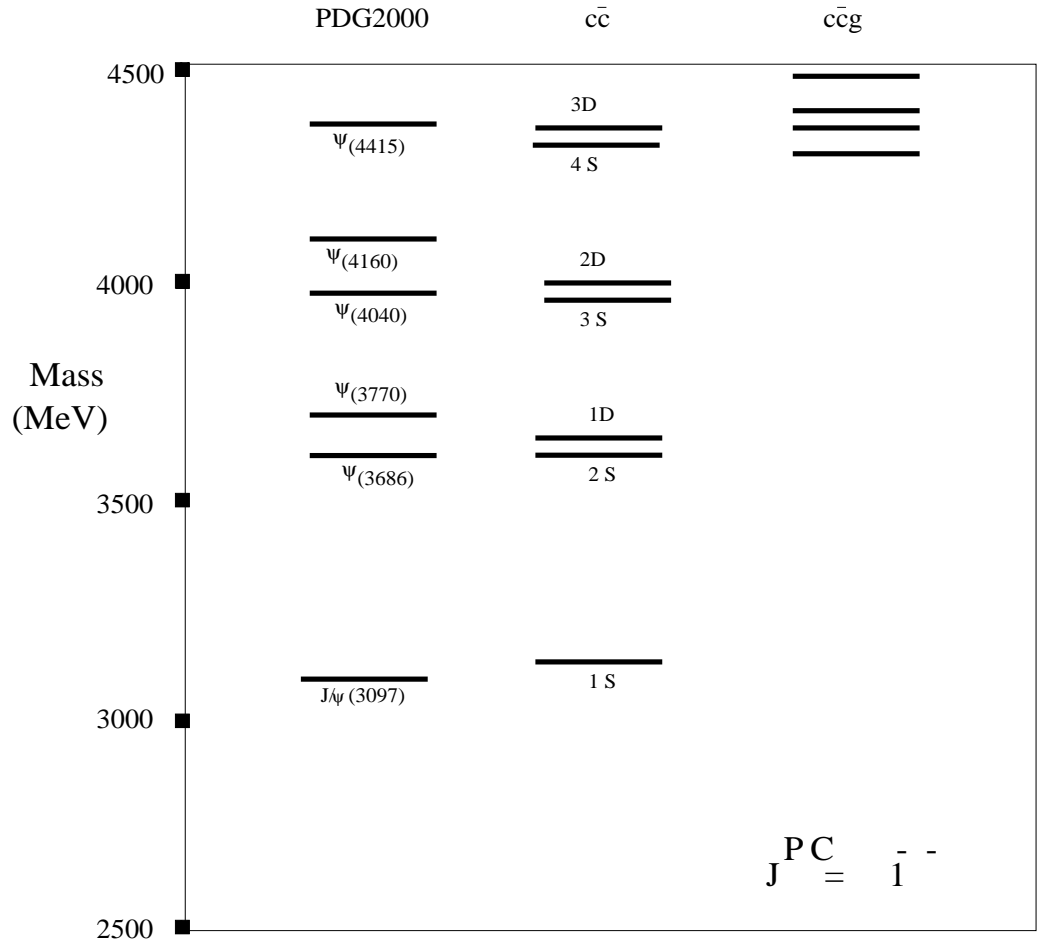,width=6in,height=5in}
\vspace{0.5in}
\caption{TDA theory
for conventional ($c\overline{c}$) and hybrid ($c\overline{c}g$)
states compared to the observed
$1^{--}$ $J/\psi$ spectrum (PDG2000) from
ref. [20]}
\label{spectrum2}
\end{figure}

Instead of solving the formidable
TDA nonlocal equations
(effectively a
12-dimensional problem in momentum
space), we evaluate the
hybrid mass
variationally
using an exponential
radial wavefunction
for each
of the two independent momentum variables.
In
the center of
momentum
frame
the  matrix elements reduce to  9-dimensional
integrals
that we evaluate
numerically using the Monte Carlo code VEGAS. We
then
perform
searches  for minima on the energy surface in the
different
angular momentum
channels. Our final results and key findings of
this
Letter are displayed in Figs. \ref{spectrum} and
\ref{spectrum2}.

Note from Fig. \ref{spectrum}  the clear agreement
between our predictions
and the lattice and flux tube results for both
light and charmed hybrid
states.
This agreement sharply contrast with the
BNL
measurements which
strongly suggests that the observed exotic states
are not
hybrids. To
confirm our result is not an artifact of the
variational method, we
have
reproduced
our conventional meson and glueball
exact TDA spectra to within
a few percent.
We also varied the least
constrained model
parameter $\alpha
_s$, from $0.2$ to $0.4$.
Related, we
even performed a more extensive RPA
variational
calculation and
took the
chiral limit ($m_u = m_d
\rightarrow
0$), finding only
miniscule change in
the hybrid mass, consistent with the
nonconservation of
the chiral
color
octet charge discussed above.  The
culmination of our
model
sensitivity study produced at most a 10 $\%$
hybrid mass
variation
indicated by the
box in Fig.  \ref{spectrum}.

Since
four quark
states
$q\overline{q}q\overline{q}$ can also have exotic quantum
numbers,
one can
make
simple estimates yielding exotic masses between 1 and 2
$GeV$
for  quarks
in color singlet configurations.
This is
consistent with a
recent unitary
quark model calculation \cite{unitary} 
which also concluded
that the
observed $1^{-+}$ states are indeed
predominately
meson-meson
resonances.

We also calculated the lightest non-exotic hybrid
(ground
state) to have
mass slightly above 2 $GeV$.  It is interesting to
note that
this state
has $J^{PC} = 1^{+-}$ in contrast to certain
heavy
hybrid,
quenched lattice ground
state results \cite{hybridlat1} which
find
near degeneracies among several 
negative
parity states.  This was
first
noted by ref. \cite{swan} who used a similar
constituent model in the
heavy
(static) quark limit.  However, the heavy quark
lattice simulations
do not
include quark spin and until this is included
the ground state
hybrid
quantum numbers, as well as related level ordering,
remains
unclear
\cite{private}.  Further, another lattice 
calculation
\cite{lacock}
studying orbital hybrid excitations concluded 
that the
ground state
quantum numbers were likely to be $1^{+-}$ in agreement with
our
work.  This study noted that the degeneracies in the static framework
would
be broken by quark spin-orbit effects which shift the exotic levels
and also
mix nonexotic and conventional $q\overline{q}$ states. They were
also able
to infer the level splitting, yielding the above conclusion,
using
propogating quarks on the lattice.  There is less, but still
some,
uncertainty in level order within our model as well since our
spin
dependent interaction is incomplete.  Although further level ordering
study
is necessary, the thrust of this work is the
exotic hybrid and
improved spin effects will not alter our conclusion
that the lightest is
above 2
$GeV$.

In Fig. \ref{spectrum2} we compare our full
model
spectrum to data for the
well studied, believed to be gluon
rich,
$1^{--}$, $J/\psi$ system to provide an explanation for the
anomalous
overpopulation of observed states \cite{PDG2000} with respect to
quark model
predictions. Whereas
previous constituent calculations, using
only $S$ waves, could only
account for 3 of
the known 6 charmonium levels,
we now predict 7
$c\overline{c}$ states in
addition to 4  $c\overline{c}g$
hybrids. Further,
ref. \cite{PDG2000} lists an
additional charmonium level
$\psi(3836)$
assigned $J^{PC}=2^{--}$ which
also agrees well with our D
wave prediction
(not shown).  Notice that by
simply including $D$ waves we
have resolved the "overpopulation" problem.  In general all of the $1^{-+}$
states, both $c\overline{c}$ and $c\overline{c}g$, will mix and a
more elaborate calculation
is in progress. However, our current result
is already sufficient to conclude
that simple level counting (density
of states) arguments will probably not
be  effective in identifying charmed
hybrid states.

Finally, we mention a novel color octet effect
leading to
an isospin splitting since it only affects the $I = 0$ states.
This is the
annihilation process depicted in Fig.
\ref{Matelement}
corresponding to
$q\overline{q} \rightarrow g \rightarrow
q\overline{q}$ for the $L_- =
0$,
spin
aligned color octet quark pair.
Octet quarkonium  is the QCD
analogue  to
ortho  positronium and the
annihilation interaction raises all
$I = 0$
light hybrid
states by roughly
$300 \
MeV$ when the
$q-\overline{q}$ spins are
aligned
($S=1$).

Summarizing, our large-scale
diagonalizations of an
effective Coulomb
gauge Hamiltonian provide a
reasonable, comprehensive
description of the meson,
lattice glueball and
lattice hybrid meson
spectra. Further, our
composite $J/\psi$
spectrum is
now also in much
better agreement with data, especially in terms of
density
of states.  It
is important to note that our quark/gluon
unified
approach
essentially
entails only  one pre-determined
dynamical
parameter. Finally, and perhaps
most significant, our
reaffirmation of
lattice  and flux tube $1^{-+}$
masses
indicates that the
recently observed exotic states are not hybrids.  Based upon
preliminary
estimates and other independent studies it is more likely
that
these
resonances are four quark states and more rigorous, higher
quark
Fock state calculations are in progress.

We thank NERSC
for
providing Cray J-90 CPU time.
F. L. E. acknowledges SURA-Jefferson Lab
for a graduate fellowship. This
work was
partially supported by grants DOE
DE-FG02-97ER41048 and
NSF
INT-9807009.

\newpage
\newpage

\end{document}